\documentclass[]{spie}  

\usepackage{amsmath,amsfonts,amssymb}
\usepackage{graphicx}
\usepackage[colorlinks=true, allcolors=blue]{hyperref}
\usepackage{aas_macros}

\title{The Dragonfly Spectral Line Mapper: Completion of the 120-lens array}

\author[a,b]{Seery Chen}
\author[c]{Deborah M. Lokhorst}
\author[d]{Imad Pasha}
\author[d]{William P. Bowman}
\author[a,b]{Qing Liu}
\author[d]{Zili Shen}
\author[c]{Aidan MacNichol}
\author[e]{Evgeni I. Malakhov}
\author[a,b]{Roberto G. Abraham}
\author[d]{Pieter van Dokkum}

\affil[a]{Dunlap Institute,
University of Toronto,
50 St. George Street, 
Toronto, ON M5S3H4, Canada}
\affil[b]{David A. Dunlap Department of Astronomy \& Astrophysics,
University of Toronto,
50 St. George Street, 
Toronto, ON M5S3H4, Canada}
\affil[c]{NRC Herzberg Astronomy \& Astrophysics Research Centre,
5071 West Saanich Road, 
Victoria, BC V9E2E7, Canada}
\affil[d]{Department of Astronomy,
Yale University,
52 Hillhouse Ave., New Haven, CT 06511, USA}
\affil[e]{New Mexico Skies, Inc., 
9 Contentment Crest, Mayhill, NM 88339, USA}

\authorinfo{Send correspondence to S.C.: seery.chen@mail.utoronto.ca}

\begin{document} 
\maketitle

\begin{abstract}
The Dragonfly Spectral Line Mapper is a mosaic telescope comprising 120 Canon telephoto lenses, based on the design of the Dragonfly Telephoto Array. With a wide field of view, and the addition of the “Dragonfly Filter-Tilter” instrumentation holding ultra narrow bandpass filters in front of each lens, the Dragonfly Spectral Line mapper is optimized for ultra low surface brightness imaging of visible wavelength line emission. The Dragonfly Spectral Line Mapper was constructed and commissioned in four phases from March 2022 to November 2023. During this time, four individual mounts of 30 lenses each were constructed and commissioned. The commissioning of the telescope included the deployment of the “Dragonfly StarChaser” which carries out image stabilization corrections in the telephoto lens, to enable hour-long exposures to be taken. In addition, we introduced new instrumentation such as a film to cover the optics to keep the filters clean. Here we describe the updated design of the complete 120-lens array, and the implementation of the instrumentation described above. Additionally, we present updated characterization of the cameras and filter transmission for the full array. Finally, we reflect on the construction and commissioning process of the complete 120-lens array Dragonfly Spectral Line Mapper, and remark on the feasibility of a larger 1000-lens array.

\end{abstract}



\section{INTRODUCTION}
\label{sec:intro}  
The Dragonfly Spectral Line Mapper is a novel narrowband imaging telescope, designed to directly image the circugalactic medium (CGM) of galaxies in the local universe \cite{chen2022, lokhorst2020-pathfinder1, lokhorst2022-pathfinder2}. The circumgalactic medium (CGM) is the gaseous medium that surrounds galaxies, containing inflows and outflows, and serving as a fuel source and regulator for star formation. The CGM plays a crucial role in galaxy formation and evolution but it is very difficult to detect due to its diffuse nature. Past studies have relied on background sources such as quasars to light up the CGM and study it in absorption\cite{tumlinson2017}, however this provides information only along pencil beam lines of sight. The CGM is expected to form structures such as filaments and fountains, with a huge variation from one galaxy to another, so looking at the CGM solely in absorption is insufficient to study the larger structure. With new instruments such as the integral field units on large telescopes such as MUSE\cite{muse2010-instrument} and KCWI\cite{keck2018-CWI}, and smaller specialized telescopes such as CHAS\cite{Melso2022-chas} and Condor\cite{Lanzetta2023-condor}, imaging the brightest parts of the CGM directly in emission lines has become a possibility. 

What the Dragonfly Spectral Line Mapper provides that other instruments do not is the the combination of all three of the following capabilities: a large field of view ($1.4^{\circ} \times 1.9^{\circ}$), narrowband imaging surface brightness sensitivity of $10^{-20}$ erg  s$^{-1 }$  cm$^{-2}$ arcsec$^{-2}$ in the H$\alpha$ emission line, and the ability to target specific recessional velocities with a resolution of $\sim$400 km/s. The large field of view is necessary for imaging the CGM out to the virial radius for local galaxies. Simulations predict that a large fraction of the cool T $\sim 10^{4}$ K CGM radiates H$\alpha$ line emission with surface brightness $\sim 10^{-20}$ erg  s$^{-1 }$  cm$^{-2}$ arcsec$^{-2}$\cite{lokhorst2019}. The ability to image gas with different line of sight velocities allows additional interpretation of data. The Dragonfly Spectral Line Mapper achieves this by combining the low surface brightness capabilities of the Dragonfly Telephoto Array \cite{2014Dragonfly} with ultranarrow-bandpass filters. By placing the filters at the aperture of each lens rather than behind the optics in a converging beam, we can use 0.8 nm bandpass filters, which is much narrower in bandwidth than those used by traditional telescopes \cite{lokhorst2020-pathfinder1}. Tilting these filters with a piece of instrumentation called the Filter-Tilter changes the central wavelength of the bandpass, allowing us to target the exact recessional velocity of each targeted galaxy. Tilting the filters from 20 to 0 degrees encompasses thousands of galaxies from the Milky Way to just beyond the Virgo Cluster.

Construction on the 120-lens Dragonfly Spectral Line Mapper was completed as of November 2023. This paper covers the hardware components that make up the telescope array, the construction process, and considerations for a future expansion of the Dragonfly Spectral Line Mapper.

\section{CURRENT CONFIGURATION OF 120-LENS ARRAY}
\label{sec:currentconfig}
\begin{figure}[h]
	\centering
    \includegraphics[width=0.95 \textwidth]{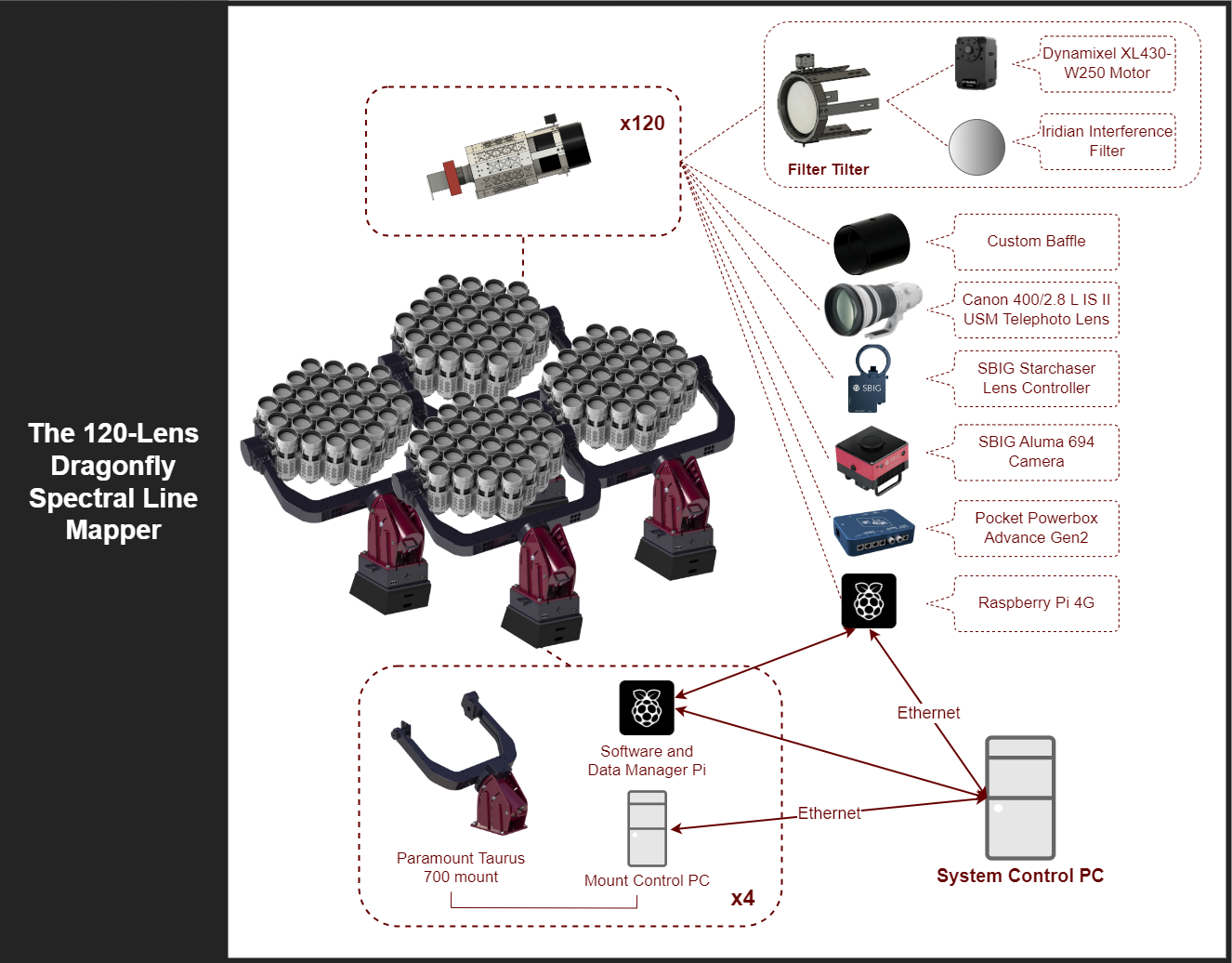}
    \caption{Updated schematic for the 120-lens Dragonfly Spectral Line Mapper. Each mount holds 30 lenses and is controlled with its own mount control PC. Each lens is a part of a fully independent subsystem with its own Filter-Tilter, baffle, lens controller, camera, Powerbox, and Raspberry Pi. One extra Raspberry Pi per mount is used to manage data and software updates. All distributed computers are controlled by the system control PC.}  
    \label{fig:120-schematic}
\end{figure}

\begin{figure}[h]
	\centering
    \includegraphics[width=1.0 \textwidth]{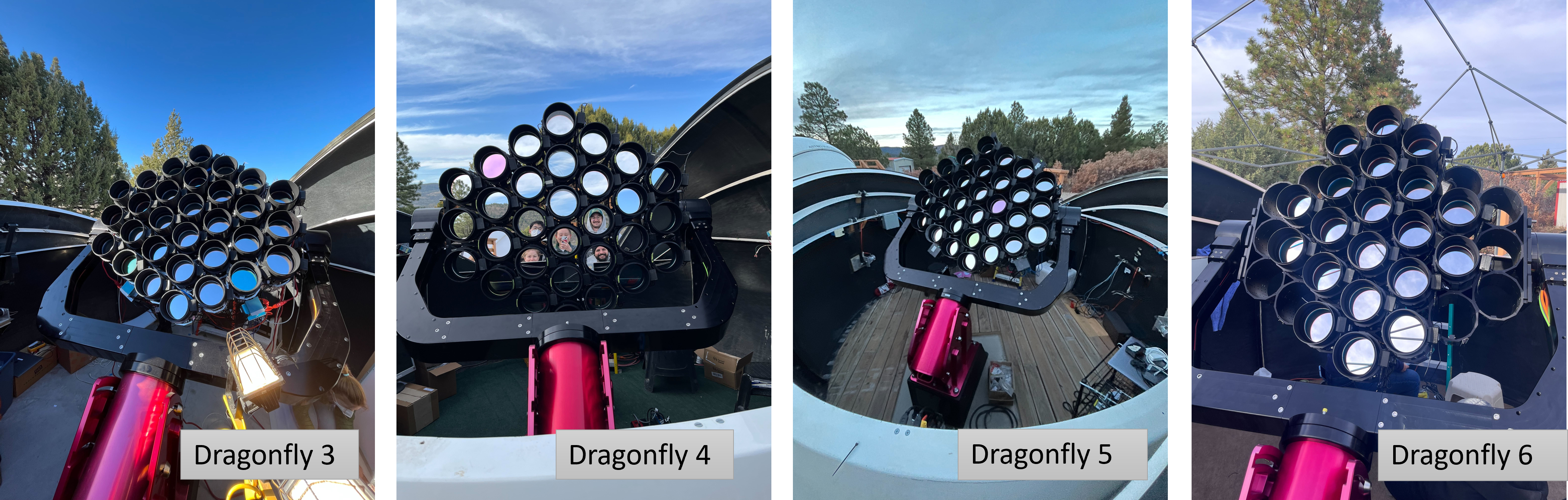}
    \caption{Images of all four Dragonfly Spectral Line Mapper (DSLM) mounts, taken from various commissioning trips at NMS. Note the first mount of DSLM is called Dragonfly 3 as the original two mounts of the broadband Dragonfly Telephoto array are Dragonfly 1 and 2.}  
    \label{fig:120lens-picture}
\end{figure}

The current configuration of the Dragonfly Spectral Line Mapper largely follows the original design plan (see Ref \citenum{chen2022}). It consists of four mounts of thirty lenses, with each lens unit functioning as a independent subsystem. See Figure \ref{fig:120-schematic} for an overview of the system and Figure \ref{fig:120lens-picture} for photos of the array. The telescope array is located at New Mexico Skies Observatories, where observing is carried out remotely, with staff on site for maintenance requests.

\subsection{Dragonfly Filter-Tilter}
The Filter-Tilter design remains unchanged. The filter is held by the inner frame of the Filter-Tilter, this inner frame is attached to the outer main frame of the Filter-Tilter through two shafts affixed with ball bearings. The DYNAMIXEL XL430-W250-T servo actuator attaches on one end of the shaft to rotate and hold the Filter-Tilter inner frame in position. For more information, see Ref \citenum{lokhorst2020-pathfinder1}, which contains specifics on the Filter-Tilter design and function, and Ref \citenum{chen2022}, which contains details about the performance of the DYNAMIXEL XL430-W250-T servo actuator.

\subsection{Filters}
The filters are same as those used on the 10-lens version of the Dragonfly Spectral Line Mapper. See Ref \citenum{chen2022} for more details on the filters and the design choices for the filter bandpasses. Here we provide a brief summary. 

 \begin{table}[ht]
\caption{Filters on the Dragonfly Spectral Line Mapper, modified from Table 3 of Ref \citenum{chen2022} to reflect the current set of filters on the instrument.The last 10 filters are currently undecided and are reserved to test potential future sets of filters. 
} 
\label{tab:DSLMFilters}
\begin{center}       
\begin{tabular}{|l|l|l|l|}
\hline
\rule[-1ex]{0pt}{3.5ex} Filter  &	Central Wavelength (nm) & Bandpass (nm) & Number on Array \\

\hline\hline
\rule[-1ex]{0pt}{3.5ex} H$\alpha$                       &	    664.7          &   0.8    &  60 \\
\hline
\rule[-1ex]{0pt}{3.5ex} H$\alpha$ Off $-$ Right         &	    611.0          &   30     &  3  \\
\hline
\rule[-1ex]{0pt}{3.5ex} H$\alpha$ Off $-$ Left          &	    705.0          &   30     &  3  \\
\hline
\rule[-1ex]{0pt}{3.5ex} \textsc{[Oiii]}                 &	    507.1          &   0.8    &  32 \\
\hline
\rule[-1ex]{0pt}{3.5ex} \textsc{[Oiii]} Off $-$ Right   &	    468.0          &   30     &  3  \\
\hline
\rule[-1ex]{0pt}{3.5ex} \textsc{[Oiii]} Off $-$ Left    &	    538.5          &   30     &  3  \\
\hline
\rule[-1ex]{0pt}{3.5ex} OH On                           &	    772.0          &   2      &  3 \\
\hline
\rule[-1ex]{0pt}{3.5ex} OH Off                          &	    768.75         &   2      &  3 \\

\hline 
\end{tabular}
\end{center}
\end{table}

\begin{figure}[h]
	\centering
    \includegraphics[width=1.0\textwidth]{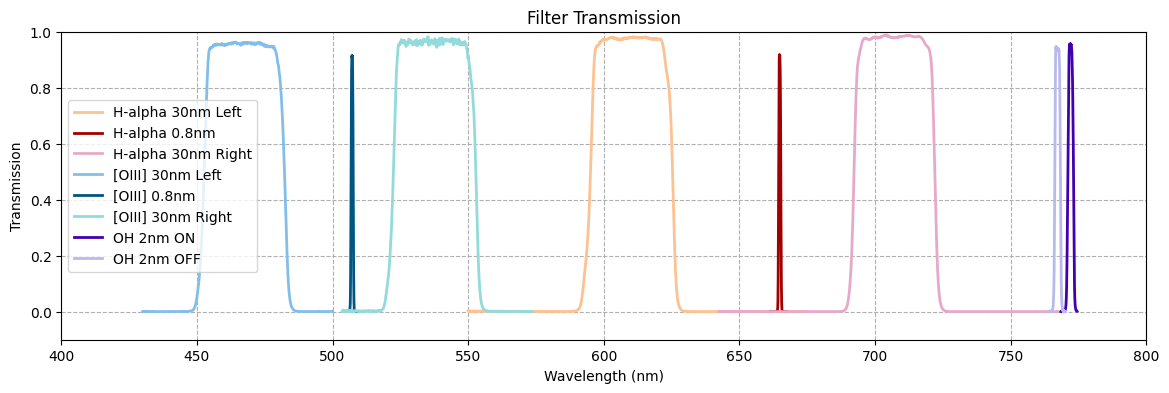}
    \caption{Transmission curves from vendor specifications for all filter types.} 
    \label{fig:allfilters}
\end{figure}

The emission lines targeted by the Dragonfly Spectral Line Mapper are H$\alpha~\lambda6563$, \textsc{[Nii]}~$\lambda6583$ and \textsc{[Oiii]}~$\lambda5007$. Due to the close proximity of the H$\alpha~\lambda6563$ and \textsc{[Nii]}~$\lambda6583$ lines, one type of filter is used to map H$\alpha$ and \textsc{[Nii]} emission. The second type of filter is used to map \textsc{[Oiii]} emission. The physical constraints of the Filter-Tilter design restrict the filter tilt angle to between $\sim -20$ to 20 degrees. This corresponds to a recessional velocity of $\sim 0 - 3900$ km/s for H$\alpha$ and \textsc{[Oiii]}~$\lambda5007$, and $\sim -1000$ to 2900 km/s for \textsc{[Nii]}. For each type of emission line filter, we have two corresponding medium-band filters with central wavelengths on either side of the emission line filter central wavelengths (H$\alpha$ Off $-$ Right/Left and \textsc{[Oiii]} Off $-$ Right/Left). These filters are used for continuum subtraction, enabling a slope in the continuum to be modeled out. The OH filter is selected to monitor OH lines in the sky spectrum in real time while observing, and an OH Off filter with a central wavelength (CWL) near the OH filter is used for continuum subtraction of the OH filter. See Figure \ref{fig:allfilters} for the transmission curves for all filter types.

\begin{figure}[h]
	\centering
    \includegraphics[width=1.0\textwidth]{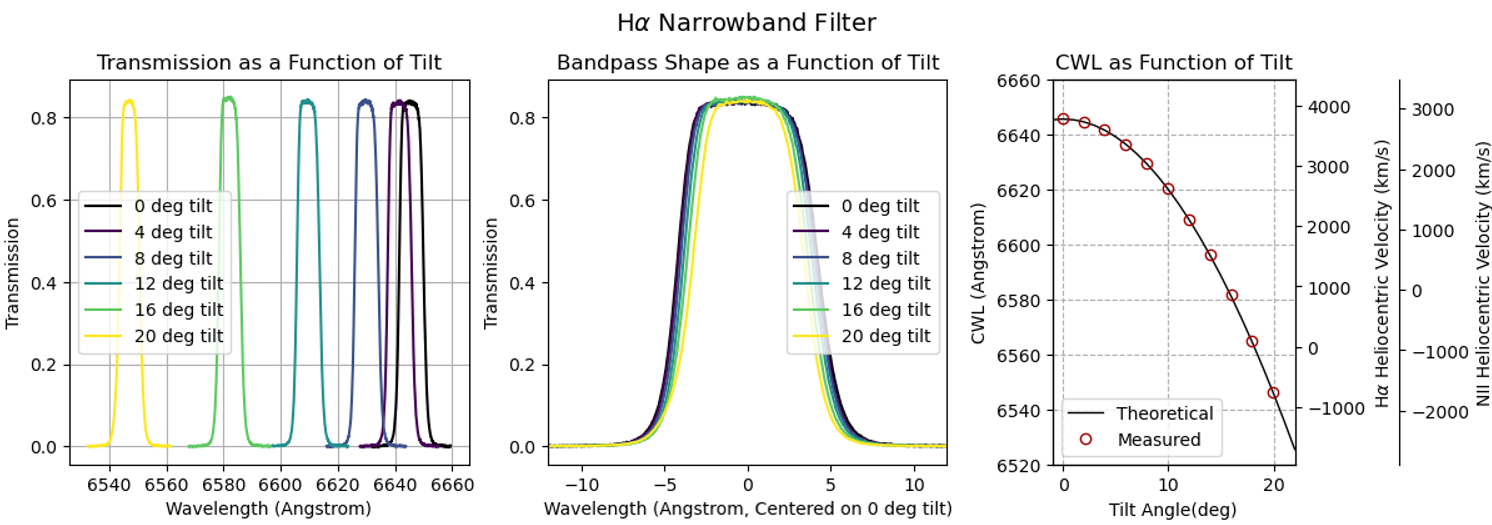}
    \caption{Left plot shows the truncated trasmission curves of the H$\alpha$ narrowband filter at varying tilt angles from 0 to 20 degrees. The middle plot shows the same transmission curves but centered on the central wavelength of the filter for comparison of the bandpass shape. Rightmost plot shows the central wavelength of the bandpass as a function of tilt angle.} 
    \label{fig:ha_bandpass}
\end{figure}

\begin{figure}[h]
	\centering
    \includegraphics[width=1.0\textwidth]{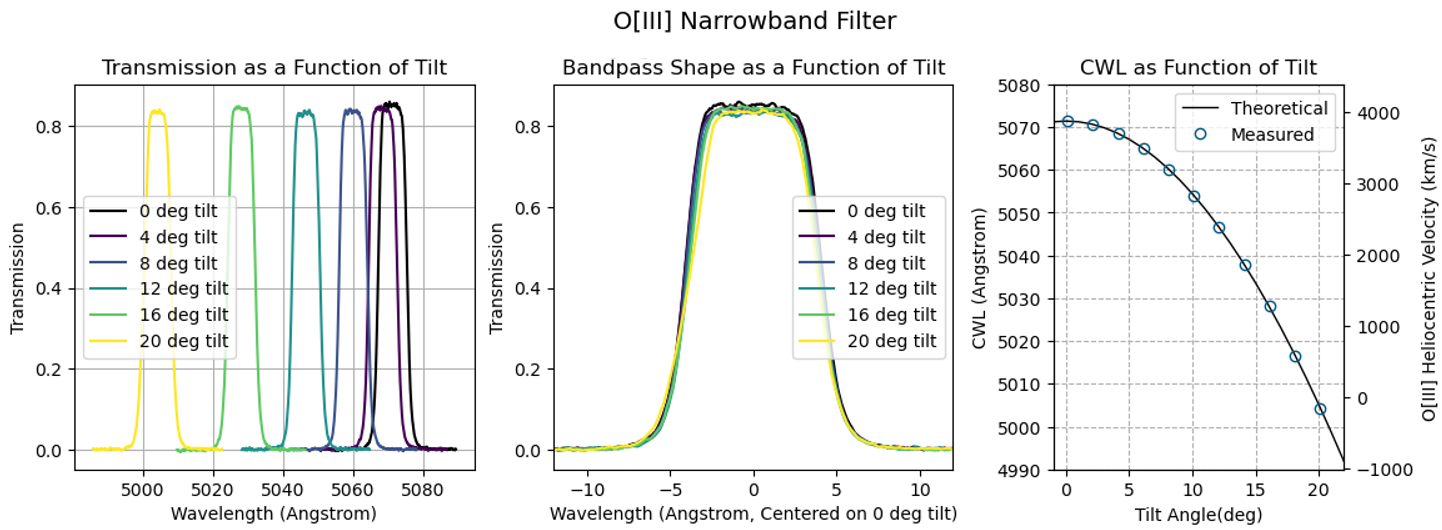}
    \caption{Same as Figure \ref{fig:ha_bandpass} but for the \textsc{[Oiii]} narrowband filter.} 
    \label{fig:oiii_bandpass}
\end{figure}

Figure \ref{fig:ha_bandpass} and Figure \ref{fig:oiii_bandpass} show enlarged views of the H$\alpha$ and \textsc{[Oiii]} narrowband filter bandpasses, respectively, as a function of filter tilt angle. The CWL $\lambda'$ of the bandpass is calculated by fitting to the following equation: 
\begin{equation}\label{supergaussian}
    T = A \exp{ (\frac{-(\lambda-\lambda')^4}{4\sigma^4}})
\end{equation}
where T is the transmission, A is a scaling factor, $\lambda$ is the wavelength, and $2(4 \log 2)^{1/4} \sigma$ is the FWHM of the bandpass. The CWL $\lambda'$ changes with tilt angle $\theta$ following 
\begin{equation}\label{n2eq}
    \lambda' = \lambda_0 \sqrt{1- (\frac{1}{n_c}\sin{\theta})^2}.
\end{equation} 
The shape of the bandpass does slightly change with increasing tilt, with the FWHM decreasing by a small amount. 

\subsection{Baffle with the Addition of Turbofilm}
\begin{figure}[h]
	\centering
    \includegraphics[width=0.45 \textwidth]{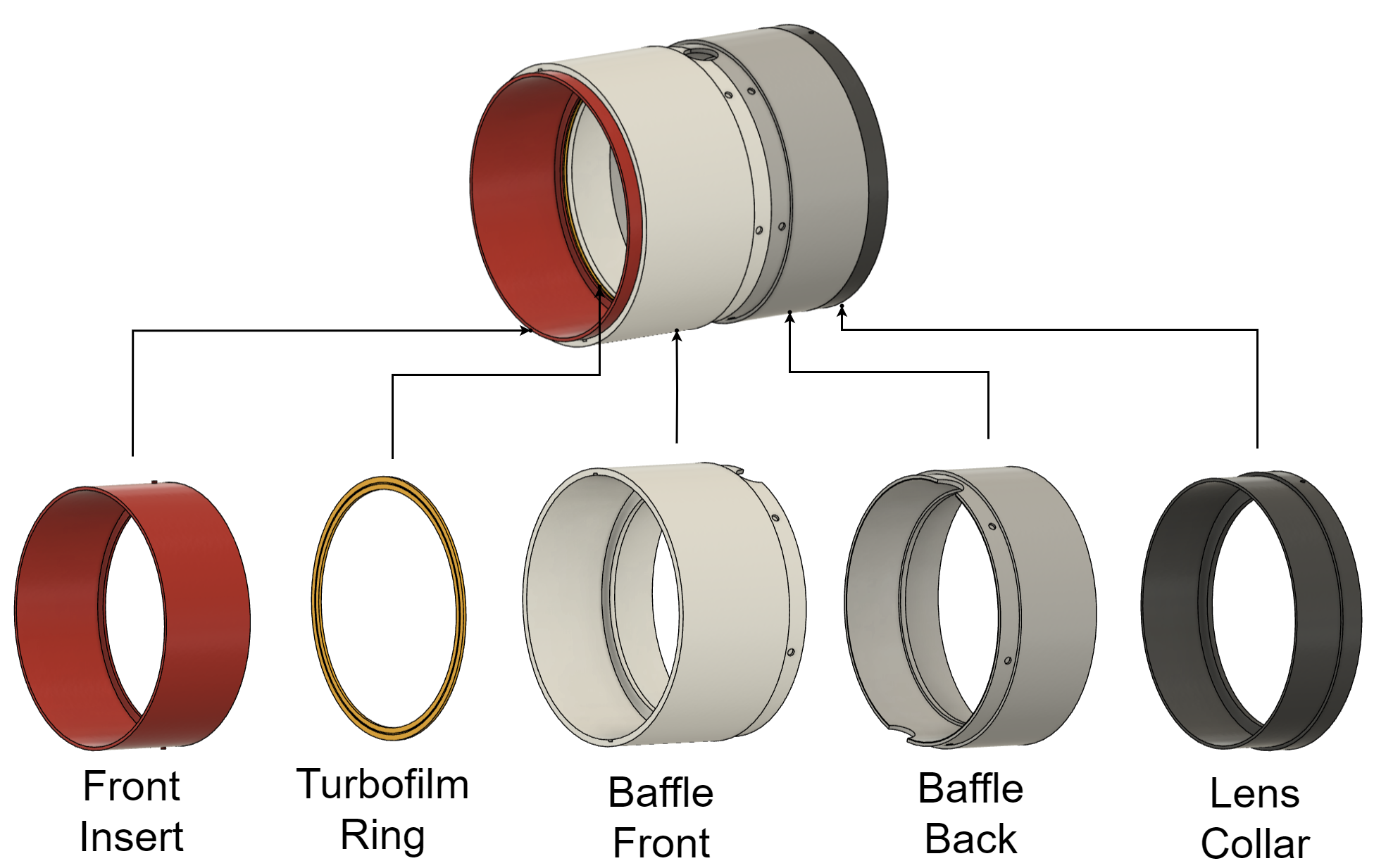}
    \caption{CAD design of the baffles used on the Dragonfly Spectral Line Mapper. Note the baffle is entirely black in colour, the different colours in this diagram are for ease of reading.}  
    \label{fig:baffle}
\end{figure}

The baffle is a custom 3D printed part, made of ABS plastic and coated with anti-reflective paint, see Figure \ref{fig:baffle}. This design remains largely the same as describe in Ref \citenum{chen2022}, with two changes in the current setup. 

The first change is the addition of Baader Turbofilm in front of the lenses. Turbofilm is a thin monomer film with an optical density of 0.1, equal to a 1/10 wave plane parallel optical window. We glue the Turbofilm on a ring, and this ring sits in front of cavity in the baffle allocated to the filter. The Turbofilm ring is stationary and covers the entirety of the front aperture, protecting the filter and lens from dust. The baffle front insert slides in with a bayonet design to lock the Turbofilm ring in place. Cleaning the filters is very time consuming and delicate task, thus having this protective film that can be changed out every 6 months saves both time and money. 

The second change to the baffle design is that the lens collar is no longer randomly orientated to increase dither between the lenses. Due to the 3D printed processes used to make the baffle, the tolerances were larger than expected, leading the lenses to be dithered more from each other than expected. 

\subsection{Lens and Lens Controller}
\begin{figure}[h]
	\centering
    \includegraphics[width=0.5 \textwidth]{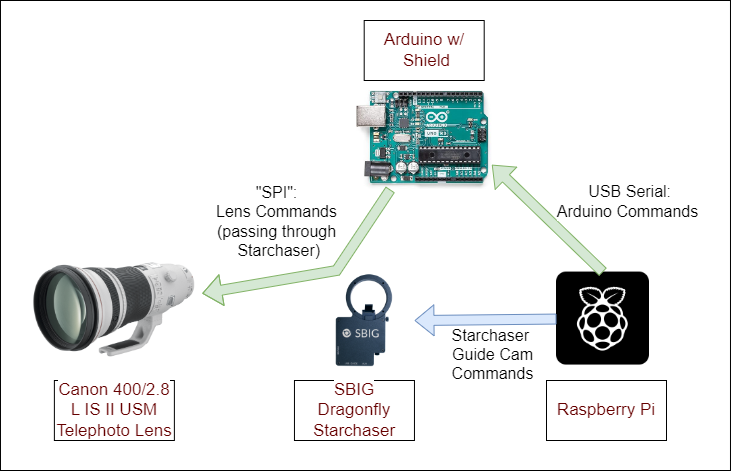}
    \caption{Schematic of the Dragonfly Starchaser system used to control the image stabilization lens and focus position of the lens. The Raspberry Pi (the control computer of each unit) sends serial commands to the Arduino, which sends command to the lens with a modified SPI protocol. This signal is transferred via the Starchaser that connects to both the Arduino and the lens.}  
    \label{fig:starchaser}
\end{figure}

The lenses used are 400 mm f/2.8 Canon IS II lenses which remain unchanged, but the devices used to send commands to the lens are new. The Birger Focuser units are replaced with a lens controller Arduino in combination with a SBIG Dragonfly Starchaser (hereafter Starchaser), see Figure \ref{fig:starchaser}. The lens controller Arduino is an Arduino Uno Rev3 microcontroller board with a custom ``Arduino shield" (a daughter board that provides extra functionality to the Arduino), and it sends commands to the lens via electrical connections provided by the Starchaser. The Starchaser is an off-axis guide camera unit with a pickoff mirror and a SONY IMX290 sensor, and it provides the physical connection between the lens and the Aluma 694 camera. The lens controller Arduino is not integrated into a microcontroller on the Starchaser as Canon's lens commands are proprietary and we were unable to share the communication protocol with a vendor. 

The reasoning behind replacing the Birger Focuser with the lens controller Arduino and Starchaser was to take advantage of the image stabilization capabilities of the lens. The Dragonfly Array mechanical design has flexure while slewing which varies on timescales of $0.5 - 1$ hr. This was not a problem for the original 48-lens Dragonfly Telephoto Array which collects 10-minute exposures. With the Dragonfly Spectral Line Mapper, though, we intend to expose for 30 minutes or longer to be sky noise limited. Inside the 400 mm f/2.8 Canon IS II lens is a small ``image stabilization" lens that can be moved in the $x y$ plane perpendicular to the optical axis $z$. Moving this lens in $xy$ moves the image on the detector plane. This IS lens was originally designed to correct for shaky images taken by hand, but would work just as well to correct for flexure at a much longer timescale. Canon uploaded custom firmware for our lenses to be able to control this image stabilization lens with their lens communication protocol. 

In lab testing for controlling the image stabilization lens and making corrections on the time scales of seconds worked well. However on-sky tests with the 0.8 nm narrow bandpass filters were unsuccessful due to the sensor in the Dragonfly Starchaser not being sensitive enough to image a sufficient number of stars to guide on. One could potentially model the flexure of the mount referencing the movement of the medium bandpass filters which let through more 30$\times$ more light, or conceive of an open loop active optics correction based on the mount pointing location. But for now, the Dragonfly Starchasers are solely used as the mechanical connection between the lens and camera, the electrical connection to the lens for focusing the lens and to keep the image stabilization lens at one place (without this command the image stabilization lens can move depending on the pointing).

\subsection{Camera}
The cameras used are the SBIG Aluma CCD694 cameras manufactured by Diffraction Limited, which are the same as those used by the 10-lens array and the 3-lens pathfinder, see Table \ref{tab:cameraspecs} for a summary of the specifications. In combination with the 400 mm f/2.8 Canon IS II lenses, the camera's pixel scale is 2.4$''$ and field of view is 1.4$^{\circ}\times$1.9$^{\circ}$.

\begin{table}[h]
\caption{Specifications for the SBIG Aluma CCD694 camera.} 
\label{tab:cameraspecs}
\begin{center}       
\begin{tabular}{|l|l|} 

\hline
\hline
\rule[-1ex]{0pt}{3.5ex}  Detector & Sony ICX-694 CCD\\
\hline
\rule[-1ex]{0pt}{3.5ex}  Pixel Size & 4.54 $\mu$m\\
\hline
\rule[-1ex]{0pt}{3.5ex}  Pixel Array & 2200x2750 pixels \\
\hline
\rule[-1ex]{0pt}{3.5ex}  Quantum efficiency (450-620nm) & 70-75\% \\
\hline
\rule[-1ex]{0pt}{3.5ex}  Quantum efficiency (750nm) & 40\%\\
\hline
\rule[-1ex]{0pt}{3.5ex}  Read noise (RMS) & 4.5 e$^{-}$\\
\hline
\rule[-1ex]{0pt}{3.5ex}  Read noise (RMS) on chip 2x2 binned & 4.5 e$^{-}$\\
\hline
\rule[-1ex]{0pt}{3.5ex}  Dark current & 0.0025-0.025e$^{-}$ pixel$^{-1}$ s$^{-1}$.\\
\hline
\rule[-1ex]{0pt}{3.5ex}  Dark current on chip 2x2 binned & 0.0094-0.086 e$^{-}$ pixel$^{-1}$ s$^{-1}$.\\
\hline
\rule[-1ex]{0pt}{3.5ex}  Full well capacity & 18,000 e$^{-}$\\
\hline
\rule[-1ex]{0pt}{3.5ex}  Gain & 0.2615 e$^{-}$ ADU$^{-1}$.\\
\hline
\rule[-1ex]{0pt}{3.5ex}  Cooling delta & 50 $^{\circ}$C \\
\hline
\hline
\end{tabular}
\end{center}
\end{table} 

\begin{figure}[h]
	\centering
    \includegraphics[width=0.95 \textwidth]{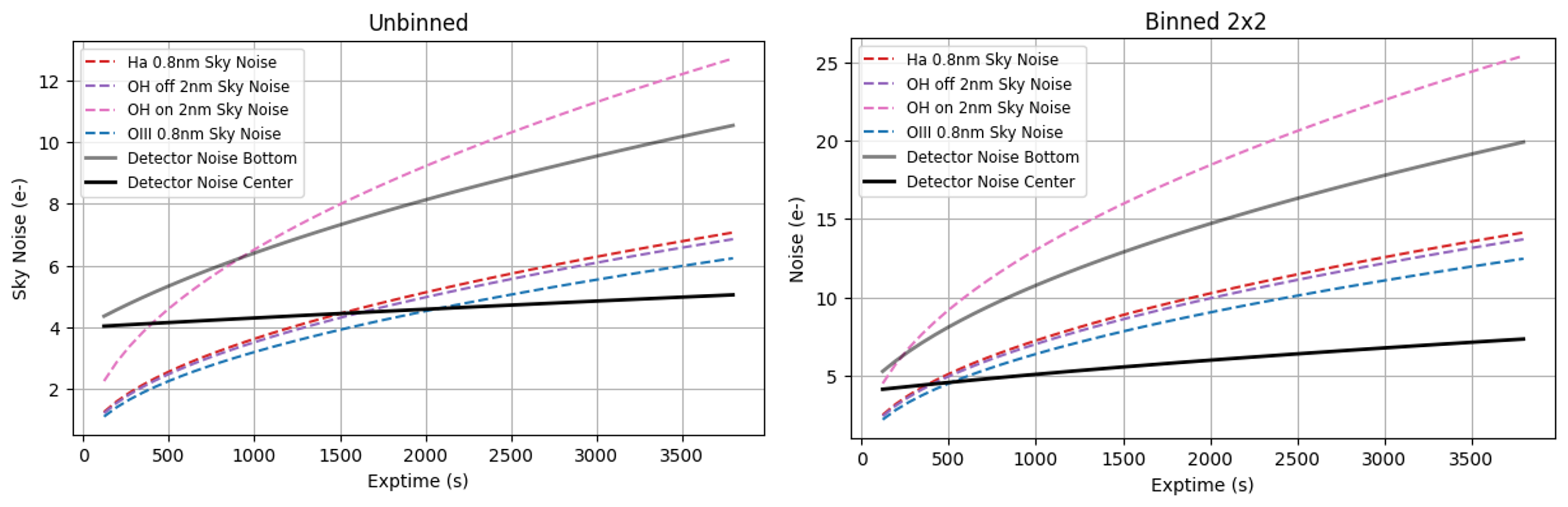}
    \caption{Sky noise in all filter bandpasses and detector noise, as a function of exposure time. Data was taken for 600, 2700, and 3600s exposures unbinned, and the square root of the sky counts was fit to a square root function. The binned sky noise is then extrapolated from the unbinned data.}  
    \label{fig:skynoise}
\end{figure}

To be sky noise limited, we need to expose for 3600s or more unbinned, and 1800s or more with 2$\times$2 on chip binning (Figure \ref{fig:skynoise}). To assess the sky counts as a function of time we take on-sky data with varying exposure times. On July 22nd and 23rd 2022, we took 600, 2700, and 3600s exposures of the galaxy M101 in the beginning of the night from 10pm to 2am, when the moon was not up and it was close to the new moon. The weather was clear. The frames were dark subtracted and flat field corrected, objects were masked and the mean sky counts were taken. The sky noise was taken to be the square root of the mean sky counts.
The detector noise was taken to be $\sqrt{\textrm{read noise} ^2 + \textrm{dark counts}}$, with the read noise being 4.3 e- and the dark counts being the dark current * time. The bottom of the detector had a dark current of 0.025 e- s$^{-1}$ pixel$^{-1}$ and the center has a dark current of 0.0025 e- s$^{-1}$ pixel$^{-1}$.
Then, we extrapolated what would be seen in binned data by multiplying the sky counts and dark counts by four, and keeping the read noise at 4.3 e-.

\begin{figure}[h]
	\centering
    \includegraphics[width=0.8 \textwidth]{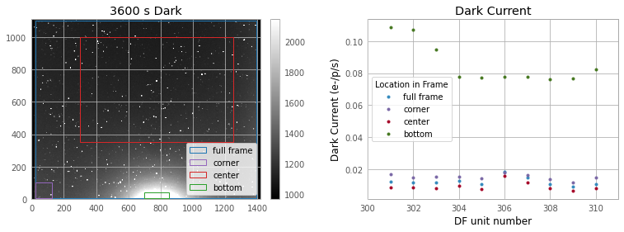}
    \caption{Left: 3600s dark iwth 2x2 on chip binning, taken by the Aluma 697 camera, with coloured boxes overlaid to show locations in the frame where the dark current was calculated. Right: the dark current calculated for each cropped section of the frame.}  
    \label{fig:2x2darkcurrent}
\end{figure}

Following the methods for measuring the dark current outlined in Ref \citenum{chen2022}, now for 2$\times$2 binned data, we cooled the detector to -20$^{\circ}$ C, collected ten darks at each of the following exposure times: 5, 30, 600, 1800, and 3600s, made a linear fit between counts and exposure time, and multiplied by the gain. We measured the 2$\times$2 binned dark current to be 0.0094 e- s$^{-1}$ pixel$^{-1}$ and 0.086 e- s$^{-1}$ pixel$^{-1}$ (center and bottom edge, respectively; Figure \ref{fig:2x2darkcurrent}).

 Following the methods for measuring readout noise outlined in Ref~\citenum{chen2022}, but now for 2$\times$2 binned data, we cooled the detector to -20$^{\circ}$ C, took ten bias frames and used the two bias subtraction method, finding the readout noise for 2$\times$2 on chip binned images is the same as the unbinned images.

\subsection{Power and Networking}
The Dragonfly Spectral Line Mapper is distributed between four domes, which are named Dragonfly 3, Dragonfly 4, Dragonfly 5, and Dragonfly 6 (counting the original Dragonfly Telephoto Array as Dragonfly 1 and Dragonfly 2). For each dome, AC power is routed through two universal power supplies, and a Digital Loggers Inc Ethernet Power Controller. Extension cords bring the power up the the DIN rail power supplies mounted on the outer lens cells. 12V DIN rail power supplies provide power to the Pegasus Pocket Powerbox Advance Gen2 (hereafter Powerbox). Each lens unit has one Powerbox to power the peripherals on that unit. The Powerbox can power cycle (turn off and on again) any of its power ports which allows individual peripherials to be power cycled without power cycling the entire array. Separate 5V DIN rail power supplies provide power to all Raspberry Pis that control each lens unit, rather than powering the Raspberry Pi with the Powerbox, as the Powerbox is a peripheral controlled by the Raspberry Pi. DIN rail Ethernet switches provide network connectivity to the Raspberry Pis.

\subsection{Computers}
Each lens unit has one Raspberry Pi computer for data acquisition and to control all the peripherals of each unit (Filter-Tilter, Starchaser and lens controller, and camera). Each mount has one PC dedicated solely to mount control.
The control PC communicates with the 120 unit Raspberry Pis and the mount PCs over Ethernet. 

A mount PC can act as a control PC for one mount and all its lenses, functioning as a fully independent subsystem (one mount unit). Due to Dragonfly 3 not being on the same sub-network as Dragonflies 4, 5, and 6, Dragonfly 3 been operating independently from the other three mounts. The lens units on Dragonflies 4, 5, and 6 are controlled by the same control PC. Once a network cable is patched over from the Dragonfly 3 dome to the other domes, the lens units on Dragonfly 3 will also be controlled by the control PC. 

The Raspberry Pi OS (a Unix-like operating system) does not seamlessly mesh with the Windows OS on the control PC, so it has been very useful to have a separate Raspberry Pi dedicated to software and data management, which is generally referred to as `pi0'. Windows only allows 26 volumes to be mounted to a computer, so we mount the data directories of the unit Raspberry Pis to pi0, then mount pi0 to the control PC to have access to all data directories of all the lens units. Pi0 is also useful for pushing software updates to the units Raspberry Pis which only have access to the internal network and no internet access.

\subsection{Mount}
The four mounts used are all Paramount Taurus 700 mounts made by Software Bisque. These are equatorial fork mounts, with a 3-axis industrial direct drive control system and on-axis absolute encoders on all three axis. They have an instrument capacity of 270 kg, an all sky pointing accuracy below 30 arcseconds RMS, and slew speed up to 30 degrees per second. The lens cell honeycomb is attached to the forks with a mounting plate that can adjust front and back to balance the load. One Orion Mini 50mm Guide Scope with a ZWO ASI120 Mini Monochrome camera sits on the top of the mount's lens cell honeycomb.

\section{CONSTRUCTION}
\label{sec:construction}

\subsection{Construction Timeline}
The Dragonfly Spectral Line Mapper was constructed in four phases from March 2022 to November 2023. 
\begin{enumerate}
    \item The first phase was from January to November 2022 with 10 lenses. In March 2022 the 10-lens iteration of DSLM completed construction, and was used to test all filter types on sky and to test new observing software. 
    \item The second phase was between November 2022 and March 2023 with 30 lenses. Two commissioning trips in November and February were required to complete the construction. This phase was critical for establishing construction methods for the 120-lens array as many hardware changes were implemented from the 10-lens to 30-lens array (including new mounts, focusers, power, and cable management).
    \item The third phase was from April 2023 to August 2023 with 60 lenses. Construction was completed with commissioning trip in April. Commissioning data was collected with the two mounts during this time.
    \item The fourth phase was from August 2023 to November 2023 with 120 lenses. Construction was completed for the last two mounts in two trips in August 2023 and November 2023 due to vendor delays. One final commissioning trip in March 2023 was needed to manage issues with the direct drive mounts. 
\end{enumerate}


\subsection{Construction Process}
Following the construction of the dome and foundation for the mount, the Paramount Taurus 700 mount was assembled. Then the lens cells were attached to the mounting plates on the mount's arms, and assembled together into the honeycomb pattern. Peripherals such as the mount guider and DIN rail power supplies at the top and bottom of the lens cell honeycomb were attached. Filters were placed into assembled Filter-Tilters, then the Filter-Tilters were attached to the lens cells, starting from the central column top to bottom, then working outwards. Then the lenses were slid into the rail mounts in the lens cells, again working top to bottom starting with the central column. After the lenses were installed, cameras with Raspberry Pis and Pegasus Powerboxes attached were attached on the back of the lenses. Finally power and network were wired to each unit. 

The observing software was installed on all the computers, the mount was balanced, tuned, polar aligned, and a pointing model was run. We evaluated the dither of the lenses and picked the most centrally pointed lens to use for a refined pointing model. Then, after the Filter-Tilters were calibrated \cite{lokhorst2022-pathfinder2}, the mount unit was ready for observing.

\section{Future directions for an expansion to the array}
\label{sec:1000lens}
The motivation for building the 120-lens Dragonfly Spectral Line Mapper is to image the circumgalactic medium. But with 1000 lenses, the Dragonfly Spectral Line Mapper could directly image the even fainter gas emission from the Intergalactic Medium, tracing the dark matter structure that spans the void between galaxies\cite{lokhorst2019}. In this section we discuss various hardware aspects to consider in an expansion of the array. 

\subsection{Cameras}
The Sony ICX-694 detector used in the Aluma694 cameras are no longer in production, however with rapid improvement to CMOS cameras in recent years, there are many CMOS cameras we can upgrade to. Cameras with the Sony IMX571 or IMX455 detectors would work well. However with the higher price for the larger chip and our Canon lenses not being able to take full advantage of that field of view, the IMX571 is a better option for our purposes. Additionally, the smaller size of the Sony IMX571 lets the Atik Apx26 camera achieve lower dark current than their camera with IMX455 chip due to the smaller sensor being easier to cool. 

\subsection{Camera to Lens connection via EF mount}
Having a tight camera to lens connection though the Canon EF mount is critical reducing systematics and data loss from flexure in the system. Due to the offbrand EF mount used in the Starchaser, and the heavy camera backend attached to each lens, there is a small amount of flexure in the system, changing the illumination on the detector depending on the gravity vector on the camera. This makes it difficult to achieve flats below the 1\% uniformity threshold. For example one flat taken with the lens right side up, and another flat taken with the lens upside down shows a gradient when dividing the two flats.  

\subsection{Direct Drive vs Geared Mounts}
Direct drive mounts require tuning the direct drive motors to be operational, and we have had many issue in tuning and maintaining this tuning of the direct drive mounts. Currently one of our four mounts is non-operational as a loosening bolt and tuning issue is being resolved. Two of four mounts have encountered tuning issues which have been resolved, where it is difficult to find a tuning model for all points in the sky, and the mounts needing to be re-tuned every few months. 

With the geared mounts on the 48-lens Dragonfly Telephoto array, this was never a problem. Fast slew speeds are not important to the Dragonfly Spectral Line mapper, thus it makes sense to use geared mounts in any future arrays.

\subsection{Assembly, Disassembly, and Maintenance}
Maintenance is a crucial aspect to the project's success, especially for a 1000-lens array that is an order of magnitude larger than the current array. Having a fully modular array that is easy to assemble and disassemble is crucial for fast repairs. Keeping the optics clean is crucial for achieving good image quality\cite{Liu2022-wideanglepsf-lensclean}. 

Currently, it very difficult to remove a Filter-Tilter without removing the lens and surrounding Filter-Tilters. This will make replacing Filter-Tilter motors or filters very difficult. To fix this, we can implement a design change where the Filter-Tilter main frame is split into two pieces, and the front including the filter and motor can be detached with a few accessible screws from the front. This would also make assembly of an entire array much easier as the legs and backside of the Filter-Tilter can be fully assembled onto the array, and lenses put into the lens cells, before the fronts with the filters are attached. 

Currently the lenses on the Dragonfly Telephoto Array are cleaned twice a year, and cleaning these 48 lenses takes a day or two. Cleaning 1000 lenses would take around a month. On the Dragonfly Spectral Line Mapper, we use Turbofilm inserts that can be replaced every 6 months. In theory this is much quicker and easier for staff to maintain. In practice the Turbofilm inserts have not been changed. Although Turbofilm is cheap and keeps the filters clean, using a removeable high transmission antireflective glass to replace turbo film that can be cleaned would make for easier maintenance. The Turbofilm inserts are difficult to remove in the current baffle design, and may introduce light scattering features in the data. Turbofilm inserts are finicky to produce as the film must be flat and have no tension on the circular frame they are glued onto. Also, they are delicate and cannot be cleaned. Thus the Turbofilm inserts are single use. Production of 1000 Turbofilm inserts on a 6 month basis is less feasible than cleaning 1000 removable glass inserts on a 6 month basis.

\section{Summary}
Construction of the 120-lens Dragonfly Spectral Line Mapper is complete as of March 2024, and it is on sky taking data. The design remains largely the same the the original design, with changes in the baffle, focuser, and the power delivery system for each unit's peripherals. Construction was completed over two years, with the array built up in stages. In a future expansion to the array where the number of lenses significantly increases, there are numerous changes to the current design to consider. Updating components such as the cameras, Filter-Tilter mechanical design, focuser (lens controller), mount, and baffle, would make for easier maintenance and better instrument performance, allowing the Dragonfly Spectral Line Mapper to potentially directly image the Intergalactic Medium.


\acknowledgments 

We are very grateful to the staff at New Mexico Skies Observatories. Their support has been crucial to this project. We are thankful for contributions from the Dunlap Institute (funded through an endowment established by the David Dunlap family and the University of Toronto) which made this research possible. This research made use of Astropy, a community-developed core Python package for Astronomy \cite{astropy:2013, astropy:2018}. We acknowledge the support of the Natural Sciences and Engineering Research Council of Canada (NSERC). Nous remercions le Conseil de recherches en sciences naturelles et en génie du Canada (CRSNG) de son soutien. We thank the Canada Foundation for Innovation (CFI) for their support.

\bibliography{report} 

\begin{thebibliography}{10}

\bibitem{chen2022}
Chen, S., Lokhorst, D.~M., Shen, J., Pasha, I., Malakhov, E.~I., Abraham, R., and van Dokkum, P., ``{The Dragonfly Spectral Line Mapper: design and first light},'' in [{\em Ground-based and Airborne Telescopes IX}{\nolinebreak\hspace{0.1em}]},  Marshall, H.~K., Spyromilio, J., and Usuda, T., eds.,  {\bf 12182},  121824E, International Society for Optics and Photonics, SPIE (2022).

\bibitem{lokhorst2020-pathfinder1}
{Lokhorst}, D.~M., {Abraham}, R.~G., {van Dokkum}, P., and {Chen}, S., ``{Wide-field ultra-narrow-bandpass imaging with the Dragonfly Telephoto Array},'' in [{\em Society of Photo-Optical Instrumentation Engineers (SPIE) Conference Series}{\nolinebreak\hspace{0.1em}]},  {\em Society of Photo-Optical Instrumentation Engineers (SPIE) Conference Series} {\bf 11445},  1144527 (Dec. 2020).

\bibitem{lokhorst2022-pathfinder2}
{Lokhorst}, D.~M., {Chen}, S., {Pasha}, I., {Shen}, J., {Malakhov}, E.~I., {Abraham}, R.~G., and {van Dokkum}, P., ``{The pathfinder Dragonfly Spectral Line Mapper: pushing the limits for ultra-low surface brightness spectroscopy},'' in [{\em Ground-based and Airborne Telescopes IX}{\nolinebreak\hspace{0.1em}]},  {Marshall}, H.~K., {Spyromilio}, J., and {Usuda}, T., eds., {\em Society of Photo-Optical Instrumentation Engineers (SPIE) Conference Series} {\bf 12182},  121821T (Aug. 2022).

\bibitem{tumlinson2017}
{Tumlinson}, J., {Peeples}, M.~S., and {Werk}, J.~K., ``{The Circumgalactic Medium},'' {\em \araa}~{\bf 55},  389--432 (Aug. 2017).

\bibitem{muse2010-instrument}
{Bacon}, R., {Accardo}, M., {Adjali}, L., {Anwand}, H., {Bauer}, S., {Biswas}, I., {Blaizot}, J., {Boudon}, D., {Brau-Nogue}, S., {Brinchmann}, J., {Caillier}, P., {Capoani}, L., {Carollo}, C.~M., {Contini}, T., {Couderc}, P., {Daguis{\'e}}, E., {Deiries}, S., {Delabre}, B., {Dreizler}, S., {Dubois}, J., {Dupieux}, M., {Dupuy}, C., {Emsellem}, E., {Fechner}, T., {Fleischmann}, A., {Fran{\c{c}}ois}, M., {Gallou}, G., {Gharsa}, T., {Glindemann}, A., {Gojak}, D., {Guiderdoni}, B., {Hansali}, G., {Hahn}, T., {Jarno}, A., {Kelz}, A., {Koehler}, C., {Kosmalski}, J., {Laurent}, F., {Le Floch}, M., {Lilly}, S.~J., {Lizon}, J.~L., {Loupias}, M., {Manescau}, A., {Monstein}, C., {Nicklas}, H., {Olaya}, J.~C., {Pares}, L., {Pasquini}, L., {P{\'e}contal-Rousset}, A., {Pell{\'o}}, R., {Petit}, C., {Popow}, E., {Reiss}, R., {Remillieux}, A., {Renault}, E., {Roth}, M., {Rupprecht}, G., {Serre}, D., {Schaye}, J., {Soucail}, G., {Steinmetz}, M., {Streicher}, O., {Stuik}, R., {Valentin}, H., {Vernet}, J., {Weilbacher}, P.,
  {Wisotzki}, L., and {Yerle}, N., ``{The MUSE second-generation VLT instrument},'' in [{\em Ground-based and Airborne Instrumentation for Astronomy III}{\nolinebreak\hspace{0.1em}]},  {McLean}, I.~S., {Ramsay}, S.~K., and {Takami}, H., eds., {\em Society of Photo-Optical Instrumentation Engineers (SPIE) Conference Series} {\bf 7735},  773508 (July 2010).

\bibitem{keck2018-CWI}
{Morrissey}, P., {Matuszewski}, M., {Martin}, D.~C., {Neill}, J.~D., {Epps}, H., {Fucik}, J., {Weber}, B., {Darvish}, B., {Adkins}, S., {Allen}, S., {Bartos}, R., {Belicki}, J., {Cabak}, J., {Callahan}, S., {Cowley}, D., {Crabill}, M., {Deich}, W., {Delecroix}, A., {Doppman}, G., {Hilyard}, D., {James}, E., {Kaye}, S., {Kokorowski}, M., {Kwok}, S., {Lanclos}, K., {Milner}, S., {Moore}, A., {O'Sullivan}, D., {Parihar}, P., {Park}, S., {Phillips}, A., {Rizzi}, L., {Rockosi}, C., {Rodriguez}, H., {Salaun}, Y., {Seaman}, K., {Sheikh}, D., {Weiss}, J., and {Zarzaca}, R., ``{The Keck Cosmic Web Imager Integral Field Spectrograph},'' {\em \apj}~{\bf 864},  93 (Sept. 2018).

\bibitem{Melso2022-chas}
{Melso}, N., {Schiminovich}, D., {Smiley}, B., {Ong}, H.~R., {Santiago}, B.~C., {Sitaram}, M., {Aleman}, I.~C., {Graber}, S., {Murillo}, M., {Rosenthal}, M., and {Stelea}, I., ``{The Circumgalactic H{\ensuremath{\alpha}} Spectrograph (CH{\ensuremath{\alpha}}S). I. Design, Engineering, and Early Commissioning},'' {\em \apj}~{\bf 941},  185 (Dec. 2022).

\bibitem{Lanzetta2023-condor}
{Lanzetta}, K.~M., {Gromoll}, S., {Shara}, M.~M., {Berg}, S., {Valls-Gabaud}, D., {Walter}, F.~M., and {Webb}, J.~K., ``{Introducing the Condor Array Telescope. I. Motivation, Configuration, and Performance},'' {\em \pasp}~{\bf 135},  015002 (Jan. 2023).

\bibitem{lokhorst2019}
{Lokhorst}, D., {Abraham}, R., {van Dokkum}, P., {Wijers}, N., and {Schaye}, J., ``{On the Detectability of Visible-wavelength Line Emission from the Local Circumgalactic and Intergalactic Medium},'' {\em \apj}~{\bf 877},  4 (May 2019).

\bibitem{2014Dragonfly}
{Abraham}, R.~G. and {van Dokkum}, P.~G., ``{Ultra-Low Surface Brightness Imaging with the Dragonfly Telephoto Array},'' {\em \pasp}~{\bf 126},  55 (Jan 2014).

\bibitem{Liu2022-wideanglepsf-lensclean}
{Liu}, Q., {Abraham}, R., {Gilhuly}, C., {van Dokkum}, P., {Martin}, P.~G., {Li}, J., {Greco}, J.~P., {Lokhorst}, D., {Chen}, S., {Danieli}, S., {Keim}, M.~A., {Merritt}, A., {Miller}, T.~B., {Pasha}, I., {Polzin}, A., {Shen}, Z., and {Zhang}, J., ``{A Method to Characterize the Wide-angle Point-Spread Function of Astronomical Images},'' {\em \apj}~{\bf 925},  219 (Feb. 2022).

\bibitem{astropy:2013}
{Astropy Collaboration}, {Robitaille}, T.~P., {Tollerud}, E.~J., {Greenfield}, P., {Droettboom}, M., {Bray}, E., {Aldcroft}, T., {Davis}, M., {Ginsburg}, A., {Price-Whelan}, A.~M., {Kerzendorf}, W.~E., {Conley}, A., {Crighton}, N., {Barbary}, K., {Muna}, D., {Ferguson}, H., {Grollier}, F., {Parikh}, M.~M., {Nair}, P.~H., {Unther}, H.~M., {Deil}, C., {Woillez}, J., {Conseil}, S., {Kramer}, R., {Turner}, J.~E.~H., {Singer}, L., {Fox}, R., {Weaver}, B.~A., {Zabalza}, V., {Edwards}, Z.~I., {Azalee Bostroem}, K., {Burke}, D.~J., {Casey}, A.~R., {Crawford}, S.~M., {Dencheva}, N., {Ely}, J., {Jenness}, T., {Labrie}, K., {Lim}, P.~L., {Pierfederici}, F., {Pontzen}, A., {Ptak}, A., {Refsdal}, B., {Servillat}, M., and {Streicher}, O., ``{Astropy: A community Python package for astronomy},'' {\em \aap}~{\bf 558},  A33 (Oct. 2013).

\bibitem{astropy:2018}
{Astropy Collaboration}, {Price-Whelan}, A.~M., {Sip{\H{o}}cz}, B.~M., {G{\"u}nther}, H.~M., {Lim}, P.~L., {Crawford}, S.~M., {Conseil}, S., {Shupe}, D.~L., {Craig}, M.~W., {Dencheva}, N., {Ginsburg}, A., {Vand erPlas}, J.~T., {Bradley}, L.~D., {P{\'e}rez-Su{\'a}rez}, D., {de Val-Borro}, M., {Aldcroft}, T.~L., {Cruz}, K.~L., {Robitaille}, T.~P., {Tollerud}, E.~J., {Ardelean}, C., {Babej}, T., {Bach}, Y.~P., {Bachetti}, M., {Bakanov}, A.~V., {Bamford}, S.~P., {Barentsen}, G., {Barmby}, P., {Baumbach}, A., {Berry}, K.~L., {Biscani}, F., {Boquien}, M., {Bostroem}, K.~A., {Bouma}, L.~G., {Brammer}, G.~B., {Bray}, E.~M., {Breytenbach}, H., {Buddelmeijer}, H., {Burke}, D.~J., {Calderone}, G., {Cano Rodr{\'\i}guez}, J.~L., {Cara}, M., {Cardoso}, J.~V.~M., {Cheedella}, S., {Copin}, Y., {Corrales}, L., {Crichton}, D., {D'Avella}, D., {Deil}, C., {Depagne}, {\'E}., {Dietrich}, J.~P., {Donath}, A., {Droettboom}, M., {Earl}, N., {Erben}, T., {Fabbro}, S., {Ferreira}, L.~A., {Finethy}, T., {Fox}, R.~T., {Garrison}, L.~H.,
  {Gibbons}, S.~L.~J., {Goldstein}, D.~A., {Gommers}, R., {Greco}, J.~P., {Greenfield}, P., {Groener}, A.~M., {Grollier}, F., {Hagen}, A., {Hirst}, P., {Homeier}, D., {Horton}, A.~J., {Hosseinzadeh}, G., {Hu}, L., {Hunkeler}, J.~S., {Ivezi{\'c}}, {\v{Z}}., {Jain}, A., {Jenness}, T., {Kanarek}, G., {Kendrew}, S., {Kern}, N.~S., {Kerzendorf}, W.~E., {Khvalko}, A., {King}, J., {Kirkby}, D., {Kulkarni}, A.~M., {Kumar}, A., {Lee}, A., {Lenz}, D., {Littlefair}, S.~P., {Ma}, Z., {Macleod}, D.~M., {Mastropietro}, M., {McCully}, C., {Montagnac}, S., {Morris}, B.~M., {Mueller}, M., {Mumford}, S.~J., {Muna}, D., {Murphy}, N.~A., {Nelson}, S., {Nguyen}, G.~H., {Ninan}, J.~P., {N{\"o}the}, M., {Ogaz}, S., {Oh}, S., {Parejko}, J.~K., {Parley}, N., {Pascual}, S., {Patil}, R., {Patil}, A.~A., {Plunkett}, A.~L., {Prochaska}, J.~X., {Rastogi}, T., {Reddy Janga}, V., {Sabater}, J., {Sakurikar}, P., {Seifert}, M., {Sherbert}, L.~E., {Sherwood-Taylor}, H., {Shih}, A.~Y., {Sick}, J., {Silbiger}, M.~T., {Singanamalla}, S.,
  {Singer}, L.~P., {Sladen}, P.~H., {Sooley}, K.~A., {Sornarajah}, S., {Streicher}, O., {Teuben}, P., {Thomas}, S.~W., {Tremblay}, G.~R., {Turner}, J.~E.~H., {Terr{\'o}n}, V., {van Kerkwijk}, M.~H., {de la Vega}, A., {Watkins}, L.~L., {Weaver}, B.~A., {Whitmore}, J.~B., {Woillez}, J., {Zabalza}, V., and {Astropy Contributors}, ``{The Astropy Project: Building an Open-science Project and Status of the v2.0 Core Package},'' {\em \aj}~{\bf 156},  123 (Sept. 2018).

\end{thebibliography}
\bibliographystyle{spiebib} 

\end{document}